\documentclass[preprint,aps,preprintnumbers,amsmath,amssymb,superscriptaddress]{revtex4}

\usepackage{graphicx}% Include figure files
\usepackage{dcolumn}% Align table columns on decimal point
\usepackage{bm}% bold math

\begin{document}

%\preprint{APS/123-QED}

\title{The negative Bogoliubov dispersion in exciton-polariton condensates}

\author{Tim Byrnes}
\affiliation{National Institute of Informatics, 2-1-2
Hitotsubashi, Chiyoda-ku, Tokyo 101-8430, Japan}

\author{Tomoyuki Horikiri}
\affiliation{National Institute of Informatics, 2-1-2
Hitotsubashi, Chiyoda-ku, Tokyo 101-8430, Japan}

\author{Natsuko Ishida}
\affiliation{Department of Information and Communication Engineering, The University of Tokyo, 7-3-1 Hongo, Bunkyo-ku, Tokyo 113-8656, Japan} 
\affiliation{National Institute of Informatics, 2-1-2
Hitotsubashi, Chiyoda-ku, Tokyo 101-8430, Japan}

\author{Michael Fraser}
\affiliation{Institute of Industrial Science,
University of Tokyo, 4-6-1 Komaba, Meguro-ku, Tokyo 153-8505,
Japan} 
\affiliation{National Institute of Informatics, 2-1-2
Hitotsubashi, Chiyoda-ku, Tokyo 101-8430, Japan}

\author{Yoshihisa Yamamoto}
\affiliation{National Institute of Informatics, 2-1-2
Hitotsubashi, Chiyoda-ku, Tokyo 101-8430, Japan}
\affiliation{E. L. Ginzton Laboratory, Stanford University, Stanford, CA 94305}

\date{\today}% It is always \today, today,
             %  but any date may be explicitly specified

\begin{abstract}
Bogoliubov's theory states that self-interaction effects in Bose-Einstein condensates produce a characteristic linear dispersion at low momenta. One of the curious features of Bogoliubov's theory is that the new quasiparticles in the system are linear combinations of creation and destruction operators of the bosons.  In exciton-polariton condensates, this gives the possibility
of directly observing the negative branch of the Bogoliubov dispersion in the photoluminescence (PL) emission.  
Here we theoretically examine the PL spectra of exciton-polariton condensates taking into account of reservoir effects.  At sufficiently high excitation densities, the negative dispersion becomes visible.  We also discuss the possibility for relaxation oscillations to occur under conditions of strong reservoir coupling.  This is found to give a secondary mechanism for making the negative branch visible. 
\end{abstract}

% PACS, the Physics and Astronomy
                             % Classification Scheme.
%\keywords{Suggested keywords}%Use showkeys class option if keyword
                              %display desired
\maketitle

Recent experimental advances have achieved the condensation of exciton-polaritons in semiconductor microcavity structures \cite{kasprzak06,deng02,deng10}. The 
short lifetime of the exciton-polaritons on the order of picoseconds means that the condensate is rather different in nature to 
a traditional atomic Bose-Einstein condensate.  The condensate exists only resulting from the replenishing of the polaritons from a reservoir of uncondensed polaritons, which is in turn populated by illumination by a laser.  Despite this difference, such condensates exhibit many of the characteristics expected in equilibrium condensates, ranging from superfluidity  \cite{amo09} to vortex formation \cite{lagoudakis08,roumpos09}.  In the work of Ref. \cite{utsunomiya08}, it was shown that 
self-interaction effects of the condensate cause the dispersion characteristics of exciton-polariton condensates follow a Bogoliubov dispersion relation. Although in the work of Ref. \cite{utsunomiya08} no negative branch of the Bogoliubov dispersion \cite{marchetti07} 
was detected, recently a four-wave mixing experiment has revealed the presence of the negative branch \cite{kohnle11}. 
The four-wave mixing experiment was originally theoretically proposed in Ref. \cite{wouters09}. The relative difficulty of
the observation of the negative branch was attributed in this work to the bright condensate emission which easily masks the 
weaker negative branch.  

In this paper we present a detailed theoretical analysis of the photoluminescence (PL) of the negative Bogoliubov branch. In contrast to the four-wave mixing approach of Ref. \cite{wouters09}, the PL is calculated directly via two-time correlation functions of the polariton equations of motion.  In particular, we incorporate the effect of the bottleneck polaritons which is known to strongly influence the dispersion of 
the polaritons  \cite{wouters07}. To this end, we first reformulate the theory of Ref.  \cite{wouters07} in a Heisenberg-Langevin formalism. The reformulation makes it clear that the theory is a modified Bogoliubov theory defining new bosonic excitations in the system. 

The polaritons are assumed to obey the Hamiltonian
\begin{align}
\label{hamiltonian}
H = & \int d^2 x \Big[ p^\dagger (\bm{x})  \left(- \frac{\hbar^2 \nabla^2}{2m} \right) p (\bm{x}) + \frac{gA}{2} p^\dagger (\bm{x}) p^\dagger (\bm{x}) p (\bm{x}) p (\bm{x}) \nonumber \\
& +\tilde{g}A p^\dagger (\bm{x}) p (\bm{x}) n_R (\bm{x})   \Big]
\end{align}
where $ p^\dagger (\bm{x}) $ is a creation operator for a polariton at position $ \bm{x} $, $ n_R (\bm{x}) $ is the number operator of the reservoir polaritons, $ m $ is the mass of a polariton, $ g $ is the interaction energy of the polaritons, and $ \tilde{g} $ is the polariton-reservoir interaction. For example, including only the exchange interaction $ g = \frac{6e^2}{4 \pi \epsilon a_B} |X|^4 \frac{a_B^2}{A} $, where $ e $ is the electronic charge, $ a_B$ is the Bohr radius, $ X $ is the excitonic Hopfield coefficient, and $ A $ is the sample area, and we use SI units \cite{ciuti01}.  The polaritons then obey the Heisenberg-Langevin equations of motion
\begin{align}
\label{equationmotion}
i \hbar \frac{d p}{dt} = [ p (\bm{x}), H] + \frac{i}{2} \left( R(n_R(\bm{x})) - \gamma \right) p (\bm{x})
\end{align}
where $  R(n_R(\bm{x})) $ is the stimulated gain coefficient \cite{scully97} of the reservoir polaritons into the condensate and $ \gamma/\hbar $ is the decay rate of the polaritons through the microcavity mirrors. Following Ref. \cite{wouters07}, we linearize equation (\ref{equationmotion})
such that it only involves terms linear in either the polariton or reservoir operators.  To achieve this goal, we expand the reservoir operator into its average and fluctuation components
\begin{align}
\label{resevoirexpansion}
n_R(\bm{x}) = n_R^0 + \frac{n_R^0}{\psi_0}\delta n (\bm{x}) 
\end{align}
where $ n_R^0 $ is the average reservoir number and $ \psi_0 $ is the amplitude of polaritons in the condensate. The prefactor of $ \delta n (\bm{x}) $ is chosen for later convenience.  Substituting (\ref{resevoirexpansion}) into (\ref{hamiltonian}) and rewriting the operators in terms of their Fourier components $ p (\bm{x}) = \frac{1}{\sqrt{A}} \sum_{\bm{k}} p_{\bm{k}} e^{i \bm{k}\bm{x}} $ and $ \delta n (\bm{x}) = \frac{1}{A} \sum_{\bm{k}} \delta n_{\bm{k}} e^{i \bm{k}\bm{x}} $, we obtain
\begin{align}
H = & \sum_{\bm{k}} ( \frac{\hbar^2 k^2}{2m} + \tilde{g} n_R^0 ) p^\dagger_{\bm{k}} p_{\bm{k}} + \frac{g}{2}  \sum_{\bm{k},\bm{k}',\bm{q}} p^\dagger_{\bm{k}+\bm{q}} p^\dagger_{\bm{k}'-\bm{q}} p_{\bm{k}'} p_{\bm{k}} \nonumber \\
& + \frac{\tilde{g} n_R^0}{\psi_0} \sum_{\bm{k},\bm{q}} p^\dagger_{\bm{k}+\bm{q}} \delta n_{\bm{q}} p_{\bm{k}} .
\end{align}
We note here that $ \delta n_{\bm{k}} $ should be interpreted as the amount of density fluctuations in the reservoir with Fourier component $\bm{k}$, and should not be confused with the number of reservoir polaritons with momentum $\bm{k}$. We now follow the same procedure to the Bogoliubov prescription to obtain a Hamiltonian that is bilinear in the variables $ (p_{\bm{k}}, p^\dagger_{-\bm{k}}, \delta n_{\bm{k}} ) $ by picking out terms in the summation which involve polariton operators with $ k=0 $, and set these to their average values $ p_{\bm{k}=0} \rightarrow  \Psi_0 = \psi_0 e^{-i\mu_T t/\hbar} $ \cite{pitaevskii03}, where $ \mu_T = \tilde{g} n_R^0 +g |\Psi_0| ^2 $ is the condensate energy.  This gives
\begin{align}
H = & \frac{g}{2} | \Psi_0|^4  + \sum_{\bm{k}} \Big[ ( \frac{\hbar^2 k^2}{2m} + \tilde{g} n_R^0 + 2 g |\Psi_0|^2) p^\dagger_{\bm{k}}  p_{\bm{k}} \nonumber \\
 + & \frac{g}{2} ( \Psi_0^2 p_{\bm{k}}^\dagger p_{-\bm{k}}^\dagger + {\Psi_0^*}^2 p_{\bm{k}} p_{-\bm{k}} ) + \frac{\tilde{g} n_R^0}{\psi_0} \delta n_{\bm{k}} ( \Psi_0 p_{\bm{k}}^\dagger +  \Psi_0^* p_{-\bm{k}} ) \Big].
\end{align}
The resulting equation of motion for the polariton operators is thus
\begin{align}
\label{pequation}
i \hbar \frac{d p_{\bm{k}}}{dt} =& \left( \frac{\hbar^2 k^2}{2m} + \tilde{g} n_R^0 + 2 g |\Psi_0|^2 \right) p_{\bm{k}} + g \Psi_0^2 p^\dagger_{-\bm{k}} \nonumber \\
& + \frac{n_R^0 \Psi_0}{\psi_0} (\tilde{g} + \frac{i}{2} R'(n_R^0)) \delta n_{\bm{k}}.
\end{align}
Meanwhile, the reservoir equation of motion obeys
\begin{align}
\label{reservoirmotion}
\frac{d n_R(\bm{x})}{dt} = G(\bm{x}) - \gamma_R n_R(\bm{x}) - R( n_R(\bm{x}) ) p^\dagger (\bm{x}) p (\bm{x}),
\end{align}
where $ G $ is the Langevin noise operator for the number operator for reservoir polaritons \cite{louisell73} and $ \gamma_R/\hbar $ is the decay rate of the reservoir polaritons. The noise operator $ G $ originates from the coupling of the reservoir modes to high energy excitations induced by the laser pump. We have neglected the diffusion of the reservoir polaritons since they have a negligible effect on the dispersion characteristics. Substituting (\ref{resevoirexpansion}) into (\ref{reservoirmotion}), expanding in Fourier space, and performing the Bogoliubov linearization we obtain
\begin{align}
\label{nequation}
\frac{d}{dt} \delta n_{\bm{k}}& = \frac{\psi_0}{n_R^0}(G_0 - \gamma_R n_R^0) \delta(k=0) - \gamma_R \delta n_{\bm{k}}
 \nonumber \\
& -  \frac{\psi_0 R(n_R^0) }{n_R^0} ( \Psi_0 p^\dagger_{\bm{k}} + \Psi_0^* p_{-\bm{k}}) - R'(n_R^0) | \Psi_0 |^2 \delta n_{\bm{k}},
\end{align}
where we have assumed a homogeneous pump $ G (\bm{x}) = G_0 $. 

The $ k \ne 0 $ components of eqns. (\ref{pequation}) and (\ref{nequation}) can be conveniently summarized in the form
\begin{align}
i \hbar \frac{d}{dt}  p_{\bm{k}}^i = \sum_{j=1}^3 M_{\bm{k}}^{ij}  p_{\bm{k}}^j
\label{matrixeqn}
\end{align}
where $ p_{\bm{k}}^i = (p_{\bm{k}} , p^\dagger_{-\bm{k}} e^{-2i\mu_T t/\hbar} ,\delta n_{\bm{k}} e^{-i\mu_T t/\hbar}) $, and
%
%\begin{widetext}
\begin{align}
M_{\bm{k}}^{ij} =
\left(
\begin{array}{ccc}
\frac{\hbar^2 k^2}{2m} + g \psi_0^2 & g \psi_0^2 &  ( \tilde{g} + \frac{i}{2} R' (n_R^0))n_R^0 \\
-g \psi_0^2 & -\frac{\hbar^2 k^2}{2m} -  g \psi_0^2 &  ( -\tilde{g} + \frac{i}{2} R' (n_R^0))n_R^0 \\
-i \frac{R(n_R^0) \psi_0^2}{n_R^0} & -i \frac{R(n_R^0) \psi_0^2}{n_R^0} & -i(\gamma_R + R'(n_R^0) \psi_0^2 )
\end{array}
\right)
\nonumber
\end{align}
%\end{widetext}
%
The matrix $ M_{\bm{k}} $ is identical to that given in Ref. \cite{wouters07}.

\begin{figure}[t]
\scalebox{0.32}{\includegraphics{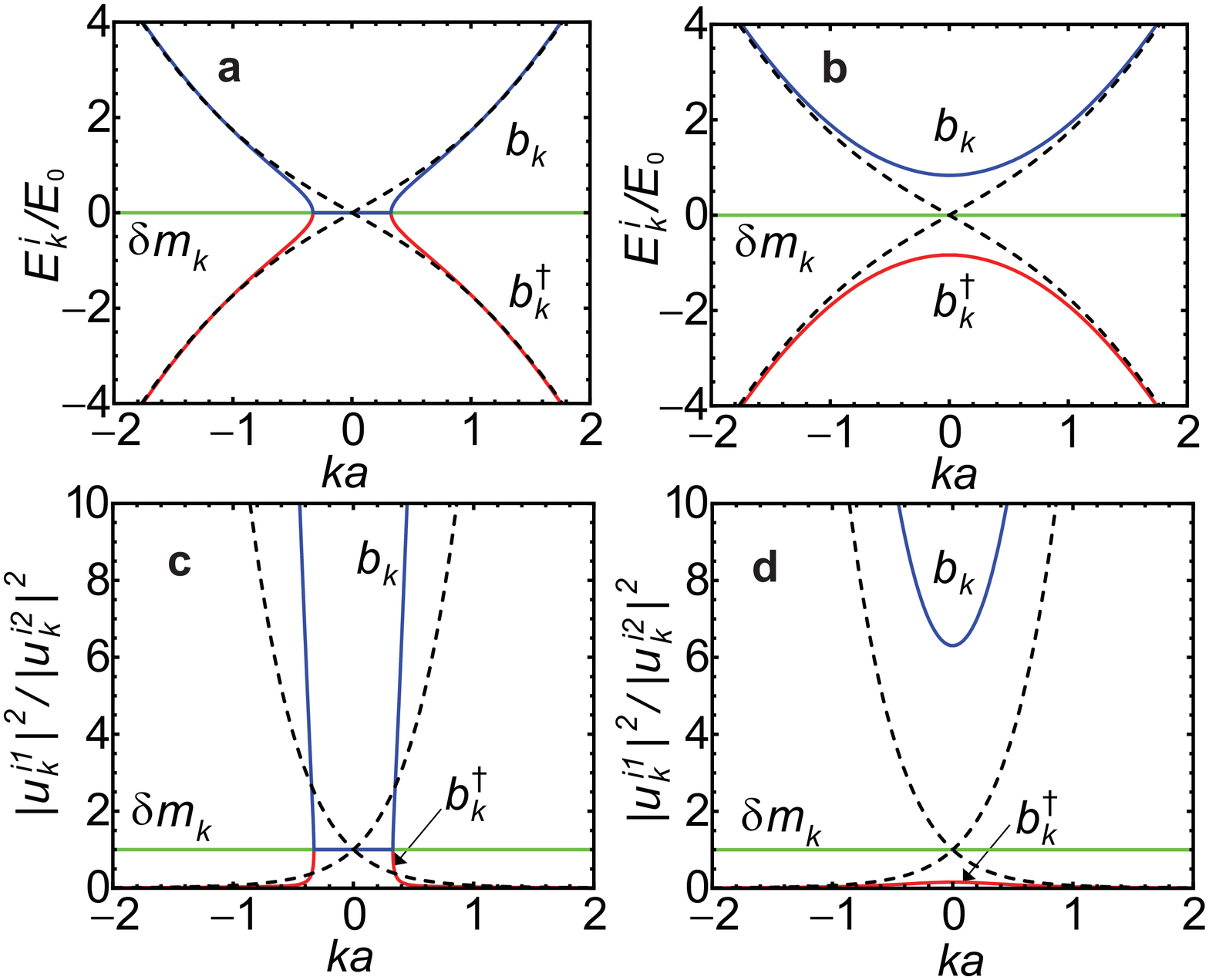}}
\caption{\label{fig1}
(Color online) (a) (b) Energy eigenvalues and (c) (d) ratio of coefficients $| u_{\bm{k}}^{i1} |^2 / | u_{\bm{k}}^{i2} |^2 $ for the eigensolutions of the matrix $ M_{\bm{k}} $ (solid lines). Dashed lines show the corresponding
values for Bogoliubov theory (no reservoir), with energy dispersion $E_{Bog} (\bm{k})/E_0 = \sqrt{(ka)^2 ( (ka)^2 +\frac{2 g\psi_0^2}{E_0})} $ and coefficients $| u_{\bm{k}}^{i1} |^2 -1/2 = | u_{\bm{k}}^{i2} |^2 +1/2 = (\frac{\hbar^2 k^2}{2m} + g \psi_0^2)/2E_{Bog} (\bm{k}) $. The energy scale is measured in units of $ E_0 = \frac{\hbar^2}{2ma^2} $, where $ a $ is the experimental length scale (e.g. in GaAs $ a = 10^{-4} $cm and $ E_0 = 0.68 $meV). Parameters used are  $ \tilde{g} n_R^0 = 1$, $ g \psi_0^2 = 1$, $ R'(n_R^0) n_R^0 = 1 $, $ R(n_R^0) = 1 $, $g= \tilde{g}$, and (a) (c) $\gamma_R= 1$ (corresponding to a flat dispersion regime) (b) (d) $\gamma_R= 0.1$ (corresponding to a relaxation oscillation regime). All parameters in units of $ E_0 $.  
}
\end{figure}

In Fig. \ref{fig1} we show the real part of the eigenvalues of $ M_{\bm{k}} $ for typical parameter values.  We identify two regimes which depend primarily on the relative magnitude of the scattering rate $ R $ and the reservoir decay rate 
$ \gamma_R $.  In Ref. \cite{wouters07}, it was generally assumed that $ \gamma_R $ was large, giving the characteristic flat dispersion regime in the vicinity of $ k = 0 $, as seen in Fig. \ref{fig1}a.  However, considering that reservoir polaritons originate from the bottleneck region \cite{tassone99} which generally have a longer lifetime than the condensate polaritons, it is possible to consider the opposite regime where $ \gamma_R $ is small and the scattering rate $ R $ is large. In such a regime (see Fig. \ref{fig1}b) the eigenspectrum shows a split dispersion at $ k=0 $, with a rounding of the Bogoliubov spectrum, such that the particles once again acquire an effective mass. This is caused by relaxation oscillations in the system, where the reservoir and the condensate repeatedly exchange population if displaced out of 
equilibrium.  A simplified description can be obtained by expanding the Gross-Pitaevskii and reservoir equations in Ref. \cite{wouters07} around their steady state values $ \psi(t) = \psi_0 + \delta \psi(t) $ and $ n_R(t) = n_R^0 + (n_R^0/\psi_0)\delta n(t) $ and ignoring interaction effects $ g = \tilde{g} = 0 $.  For small $ \bm{k} $ we obtain
\begin{align}
\hbar \frac{d}{dt}
\left(
\begin{array}{c}
\delta \psi \\
\delta n
\end{array}
\right)
 =
\left(
\begin{array}{cc}
0 &     \frac{1}{2} R' (n_R^0)n_R^0 \\
- \frac{2 R(n_R^0) \psi_0^2}{n_R^0} &  -\gamma_R - R'(n_R^0) \psi_0^2 
\end{array}
\right)
\left(
\begin{array}{c}
\delta \psi \\
\delta n
\end{array}
\right)
\nonumber 
\end{align}
The imaginary part of the eigenvalues gives the oscillation frequency, which is 
\begin{align}
\omega = \frac{1}{\hbar}\sqrt{ R' (n_R^0)R(n_R^0) \psi_0^2-(\gamma_R + R'(n_R^0) \psi_0^2 )^2/4 } .
\label{relaxationfreq}
\end{align}
In the limit of small scattering $ R(n_R) $ or large $ \gamma_R $ the frequency becomes pure imaginary, indicating that only damping occurs. 

The eigenvalues of the matrix $ M_{\bm{k}} $ give the new collective excitations of the system. Specifically, new quasiparticle operators
\begin{align}
\label{newoperator}
b_{\bm{k}}^i = \sum_{j=1}^3 u_{\bm{k}}^{ij} p_{\bm{k}}^j
\end{align}
may be defined where $ u_{\bm{k}}^{ij} $ are the eigenvectors of $ M_{\bm{k}} $.  The matrix $ M_{\bm{k}} $ allows for two types of solutions, either dispersionless or bosonic, defined as whether the commutation relation
\begin{align}
\label{commutationequation}
[b_{\bm{k}}^i ,{b_{\bm{k}}^i}^\dagger] =  | u_{\bm{k}}^{i1} |^2 - | u_{\bm{k}}^{i2} |^2 
\end{align}
is zero or non-zero respectively. Here we used the identities $ \delta n_{\bm{k}}^\dagger = \delta n_{-\bm{k}} $ and $ [ \delta n_{\bm{k}},  \delta n_{-\bm{k}}] = 0 $. Operators with non-zero commutators may be normalized to $\pm 1 $, defining new bosonic modes of the system. From Fig. \ref{fig1}c we see that no bosonic solutions exist in the flat dispersion regime, since all eigenvectors satisfy $ | u_{\bm{k}}^{i1} |^2 = | u_{\bm{k}}^{i2} |^2 $, which corresponds to $ [b_{\bm{k}}^i ,{b_{\bm{k}}^i}^\dagger] = 0 $ from (\ref{commutationequation}).  In the 
the regime beyond the flat dispersion there is always one dispersionless solution with $ | u_{\bm{k}}^{i1} |^2 = | u_{\bm{k}}^{i2} |^2 $ corresponding to a renormalized density fluctuation solution.  The solution with $ | u_{\bm{k}}^{i1} |^2 > | u_{\bm{k}}^{i2} |^2 $ corresponds to a solution with $ [b_{\bm{k}} ,{b_{\bm{k}}}^\dagger] = 1 $, which is a 
new ``dissipative'' Bogoliubov destruction operator.  The solution with $ | u_{\bm{k}}^{i1} |^2 < | u_{\bm{k}}^{i2} |^2 $ meanwhile 
corresponds to $ [b^\dagger_{-\bm{k}} ,b_{-\bm{k}}] = -1 $, the dissipative Bogoliubov creation operator (see also Fig. \ref{fig1}d). Putting in the correct time dependences, outside the flat dispersion regime we may associate the solutions to be $ b_{\bm{k}}^i \equiv (b_{\bm{k}} , b^\dagger_{-\bm{k}} e^{-2i\mu_T t/\hbar} ,\delta m_{\bm{k}} e^{-i\mu_T t/\hbar}) $, corresponding to the positive, negative, and dispersionless branches respectively.  Since (\ref{newoperator}) appears to admix both bosonic and number operators one may wonder how to interpret such an operator. The clearest interpretation is that this is a displaced Bogoliubov operator 
$ b_{\bm{k}} = b_{\bm{k}}' + u_{\bm{k}}^{13} \delta n_{\bm{k}} $, where the 
first two components form the boson operator and the reservoir creates displacements in the vacuum from this state.  In this 
case the amount of reservoir fluctuations of wavelength $ k $ displaces the vacuum for Bogoliubov particles $ b_{\bm{k}}'$ at momentum $ k $.

To calculate the PL spectrum we apply the methods presented in Ref. \cite{ciuti01}.  The PL spectrum is given by
\begin{align}
& PL(\bm{k},\omega)  \propto |C_{\bm{k}}|^2 \mbox{Re} [ \langle p_{\bm{k}}^\dagger (t=0) \tilde{p}_{\bm{k}} (\omega)  \rangle ] \nonumber \\
& = |C_{\bm{k}}|^2 \mbox{Re} [ \sum_{i j} ({\bar{u}}_{\bm{k}}^{1i})^{*} \bar{u}_{\bm{k}}^{1j} \langle {b_{\bm{k}}^i}^\dagger (t=0) \tilde{b}_{\bm{k}}^j (\omega)  \rangle]
\label{plexpansion}
\end{align}
where $ \bar{u}_{\bm{k}} $ is the inverse of $ u_{\bm{k}} $ (normalized to satisfy bosonic commutation relations), $ C_{\bm{k}} $ is the Hopfield coefficient for the 
photonic component of the polaritons, and tildes denote time Fourier transformed variables. 
In the diagonal basis the equations of motion (\ref{matrixeqn}) are
\begin{align}
i \hbar \frac{d b_{\bm{k}}^i}{dt} & = E^i_{\bm{k}} b_{\bm{k}}^i + i F_{\bm{k}}^i 
\label{bequation}
\end{align}
where we have added a Langevin noise operator $ F_{\bm{k}}^i = (F_{\bm{k}}, F_{-\bm{k}}^\dagger, 0) $ to account for a thermal population of dissipative Bogoliubov particles \cite{pitaevskii03}, and $ E^i_{\bm{k}} $ are the eigenvalues of $ M_{\bm{k}} $. The 
noise operator is assumed to obey correlations of the form \cite{scully97}
\begin{align}
\langle b_{\bm{k}}^i F_{\bm{k}}^\dagger \rangle & = \delta_{i1} \Gamma_{\bm{k}}  
n_{\bm{k}}^{\mbox{\tiny th}} 
\label{reservoircorrelations}
\end{align}
where $ n_{\bm{k}}^{\mbox{\tiny th}} = 1/(e^{ \epsilon_{\bm{k}}/k_B T} -1 ) $, 
$ \epsilon_{\bm{k}} = \mbox{Re} [ E^1_{\bm{k}} ] = -  \mbox{Re} [ E^2_{\bm{k}} ] $, $ \Gamma_{\bm{k}} = \mbox{Im} [E^1_{\bm{k}}] = \mbox{Im} [E^2_{\bm{k}}] $, $ T $ is the temperature, and $ \delta_{ij} $ is the Kronecker delta. This ensures that the dissipative Bogoliubov particles obey thermal statistics $ \langle b_{\bm{k}}^\dagger b_{\bm{k}} \rangle = n_{\bm{k}}^{\mbox{\tiny th}} $, with other off-diagonal correlations vanishing. The choice of (\ref{reservoircorrelations}) 
is chosen primarily because it is the simplest choice, and is possible to generalize to other distributions \cite{sarchi08}. 
 
\begin{figure}
\scalebox{0.32}{\includegraphics{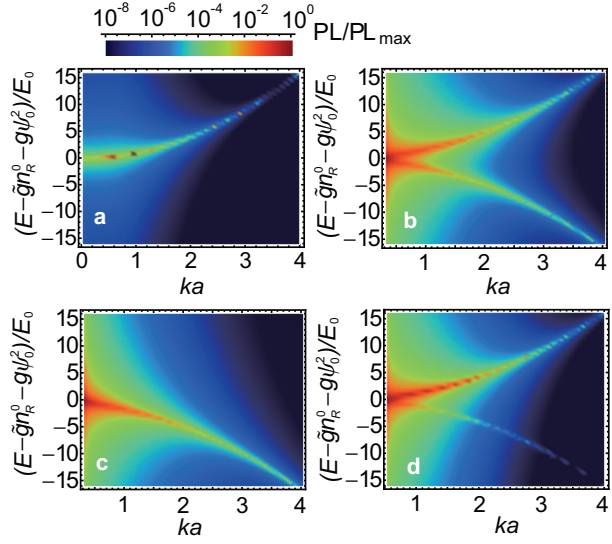}}
\caption{\label{fig2}
Photoluminescence of the excitation spectrum of exciton-polariton condensates for various parameters. Regimes of (a) just above condensation threshold (b) thermally depleted high density (c) high density and low temperature (d) zero interactions are shown.  Parameters used are $ \tilde{g} n_R^0 = 1$,  $ R'(n_R^0) n_R^0 = 1 $, $ R(n_R^0) = 1 $, $g= \tilde{g}$,  $\gamma_R= 1$, and (a) $ \psi_0^2 =0.001 n_R^0$, $ k_B T = 2 $, (b) $ \psi_0^2 =n_R^0$, $ k_B T = 2 $,
(c)  $ \psi_0^2 =n_R^0$, $ k_B T = 0.1 $, (d) $ g = \tilde{g} = 0 $,  $ \psi_0^2 =n_R^0$, $ k_B T = 2 $.  A chemical potential of $ \mu = -1 $ was assumed in the thermal distribution in order to account for finite size effects of the condensate.
All parameters in units of $ E_0 $.  Zero detuning and a Rabi splitting of $10 E_0 $ was assumed to calculate the Hopfield coefficient $ C_{\bm{k}} $. }
\end{figure}

After a Fourier transform of (\ref{bequation}) and inserting the correlations we obtain the expression for the PL
\begin{align}
PL(\bm{k},\omega) \propto |C_{\bm{k}}|^2 \mbox{Re} \left[
\frac{i | \bar{u}_{\bm{k}}^{11} |^2 n_{\bm{k}}^{\mbox{\tiny th}}}{\hbar \omega - \epsilon_{\bm{k}} - i \Gamma_{\bm{k}}  } 
+ \frac{i | \bar{u}_{\bm{k}}^{12} |^2 (n_{\bm{k}}^{\mbox{\tiny th}} +1)}{\hbar \omega + \epsilon_{\bm{k}} - i \Gamma_{\bm{k}}} \right] .
\label{plfinalequation}
\end{align}
Here the first term corresponds to the positive dispersion, weighted by 
the thermal population and the dissipative Bogoliubov coefficient, while the second term is the negative dispersion, which only appears when there
is appreciable Bogoliubov mixing between the $ p_{\bm{k}} $ and $ p_{-\bm{k}}^\dagger $ operators.  In Fig. \ref{fig2}a and \ref{fig2}b we plot the PL spectrum corresponding to low and high density regimes respectively. At low density we see that only the positive branch is visible. Here there is negligible mixing between the components (\ref{newoperator}) due to the small off-diagonal components of $ M_{\bm{k}} $. The positive branch is populated via the 
thermal noise field $ F_{\bm{k}} $, while the negative branch remains dark.  At sufficiently high density, the off-diagonal components of $ M_{\bm{k}} $ become large enough such that there is some mixing of the components (\ref{newoperator}).  This results in a visible negative dispersion in the PL spectrum. A simple criterion for when the negative branch becomes visible may be derived. As may be seen by inspection of the Bogoliubov expression for the factor $ | \bar{u}_{\bm{k}}^{12} |^2 = (\frac{\hbar^2 k^2}{2m} + g \psi_0^2)/2E_{Bog} (\bm{k}) -1/2 $, the 
negative component is only appreciable for momenta $ \frac{\hbar^2 k^2}{2m} < g \psi_0^2 $. Due to the finite linewidth of the dispersion, the negative dispersion only becomes resolvable beyond momenta $ \frac{\hbar^2 k^2}{2m} > \Gamma_{\bm{k}}  $.  Thus the 
negative dispersion only becomes visible for densities exceeding the criterion $ g \psi_0^2 \sim \Gamma_{\bm{k}} $.   

It is interesting that for low temperatures where $ n_{\bm{k}}^{\mbox{\tiny th}} $ is small, only the negative excitation branch is visible in the PL spectrum (Fig. \ref{fig2}c). The reason for this can be understood by the following argument.  According to the inverse relation of (\ref{newoperator}), the loss of a polariton out of the system (which is the basic process underlying the PL) is described by the operator
\begin{align}
p_{\bm{k}} = \bar{u}_{\bm{k}}^{11} b_{\bm{k}} + \bar{u}_{\bm{k}}^{12} b^\dagger_{-\bm{k}} e^{-2i\mu_T t/\hbar} +
\bar{u}_{\bm{k}}^{13} \delta m_{\bm{k}} e^{-i\mu_T t/\hbar} .
\end{align}
The energy change of the system associated with the first term is $ \mu_T + \epsilon_{\bm{k}}  $, which is nothing but the standard mechanism for the PL with a positive dispersion. With no thermal population of 
dissipative Bogoliubov particles, the first term automatically gives zero,
thus the positive dispersion remains dark.  The second term is associated with the gain of a dissipative Bogoliubov particle and the loss of two condensate particles, which has an energy change of $ 2 \mu_T - (  \mu_T + \epsilon_{\bm{k}} ) = \mu_T - \epsilon_{\bm{k}} $.
Unlike the first term where a dissipative Bogoliubov particle needs to be originally present, the second term does not require this condition and occurs regardless of the initial population. We note that a similar effect has been observed in atom lasers \cite{japha99}. The third term causes a loss of a condensate particle with an energy change of $ \mu_T  $ independent of $ \bm{k} $, giving a dispersionless spectrum.   This can in principle give rise to a flat PL emission as seen in Fig. \ref{fig1} if there are strong enough density fluctuations in the reservoir, although we do not assume that this is typically the case in current experimental systems.

There is another mechanism for making the negative branch visible, which is in a regime where reservoir scattering is strong.  To see this effect, we show in Fig. \ref{fig2}d the PL spectrum for the illustrative case of zero interactions $ g = \tilde{g} = 0 $. We again see that the negative branch is visible along the eigenvalues of $ M_{\bm{k}} $. In this case the negative branch is visible not because of self-interactions, but due to mediation via the reservoir mode. 
From our simulation results we find that the negative branch becomes visible for the same condition as that given to observe
relaxation oscillation (i.e. that (\ref{relaxationfreq}) is real). The mechanism for this is that due to the relaxation oscillations there is an effective coupling of the $ p_{\bm{k}} $ and $ p_{-\bm{k}}^\dagger $ operators mediated by the 
$ \delta n_{\bm{k}} $ operator. In practice both the self-interactions and the reservoir effect are likely to contribute to making the negative dispersion visible. 

In summary, we have calculated the PL spectrum of the excitations of exciton-polariton condensates.  At sufficiently high densities such that $ g \psi_0^2 \sim \Gamma_{\bm{k}} $, the negative branch of the Bogoliubov spectrum becomes visible. 
The negative dispersion should be visible even if no thermal population is present, due to the nature of the PL emission being
associated with the loss of a polariton from the system. In the regime of long reservoir decay times, a regime of relaxation
oscillations was identified, which was also found to be able to illuminate the negative dispersoin.  One question which remains is why a four-wave mixing experiment was required in order to see the negative dispersion in Ref. \cite{kohnle11}, instead of spontaneously.  One explanation is that the bright emission from the condensate itself may be masking the negative branch, and only the faint tail at high momenta can be accessed \cite{wouters09}. Another possibility is the reduction of the condensate self-interactions at higher density due to reduction of the Bohr radius \cite{byrnes10,kamide10}. This may contribute to make the negative branch weaker than expected due to the suppression of Bogoliubov mixing.

We thank J. Keeling for discussions.  
This work is supported by the Special Coordination Funds for Promoting Science and Technology, the FIRST program for JSPS, Navy/SPAWAR Grant N66001-09-1-2024, MEXT, and NICT.

%%%%%%%%%%%%%%%%%%%%%%%%%%%%%%%%

\end{document}